# On the Ferroelectric Polarization Switching of Hafnium Zirconium Oxide in Ferroelectric/Dielectric Stack

Mengwei Si, Xiao Lyu, and Peide D. Ye*

*School of Electrical and Computer Engineering and Birck Nanotechnology Center, Purdue University, West Lafayette, Indiana 47907, United States*

* Address correspondence to: yep@purdue.edu (P.D.Y.)

ABSTRACT

The ferroelectric polarization switching in ferroelectric hafnium zirconium oxide ($Hf_{0.5}Zr_{0.5}O_2$, HZO) in the HZO/$Al_2O_3$ ferroelectric/dielectric stack is investigated systematically by capacitance-voltage and polarization-voltage measurements. The thickness of dielectric layer is found to have a determinant impact on the ferroelectric polarization switching of ferroelectric HZO. A suppression of ferroelectricity is observed with thick dielectric layer. In the gate stacks with thin dielectric layers, a full polarization switching of the ferroelectric layer is found possible by the proposed leakage-current-assist mechanism through the ultrathin dielectric layer. Theoretical simulation results agree well with experimental data. This work clarifies some of the critical parts of the long-standing confusions and debating related to negative capacitance field-effect transistors (NC-FETs) concepts and experiments.

## Introduction

A ferroelectric material has two stable polarization states with different directions, which are switchable by the external electric field, and thus is extensively explored for non-volatile memory applications. Using ferroelectric field-effect transistors (Fe-FETs) as FET-type ferroelectric memory is a promising ferroelectric memory architecture, because of its high density and non-destructive readout.[1–4] Recently, using a ferroelectric-gated transistor as a negative capacitance field-effect transistor (NC-FET) has attracted tremendous attention as a novel steep-slope device.[5–9] In both Fe-FET and NC-FET, a ferroelectric (FE) insulator and linear dielectric (DE) insulator bilayer stack[10-17] is applied as the gate structure. The necessity of such a linear DE layer is because an interfacial oxide layer between semiconductor channel and FE insulator is required to improve the ferroelectric/semiconductor interface and meanwhile provide sufficient capacitance matching if quasi-static negative capacitance (QSNC) concept is applied for the development of NC-FETs.[5] QSNC definition was introduced to distinguish the stabilized NC effect and transient NC effect.[18] One important fact, which has been overlooked in the past several years, is that the FE/DE stack capacitor is fundamentally different from a FE capacitor and DE capacitor in series.[18,19] Therefore, the complete understanding of the impact of DE layer on the ferroelectric properties of a FE/DE stack is crucial to study the ferroelectric switching mechanism in Fe-FETs and NC-FETs.

Ferroelectric hafnium oxide, such as hafnium zirconium oxide ($Hf_{0.5}Zr_{0.5}O_2$, HZO), has been recently discovered as an ultrathin CMOS compatible high performance ferroelectric insulator.[20,21] Therefore, HZO is chosen as the FE insulator for this study and $Al_2O_3$ is chosen as the linear DE insulator to study the ferroelectric polarization switching in the FE/DE stack. As is well-known that ferroelectric HZO has a thickness-dependent remnant polarization ($P_r$) at about 10-30 μC/cm$^2$.[2,22,23] However, a conventional dielectric insulator cannot support such a large charge density. For example, $Al_2O_3$ has a typical dielectric constant of 8 and a breakdown electric field less than 1 V/nm.[24] The calculated charge density at breakdown electric field is about 7 μC/cm$^2$, which is the maximum charge density ($Q_{max}$) for ideal $Al_2O_3$ to be able to support without breakdown. Note that in reality, this value for $Q_{max}$ is much smaller. As can be seen that, even in ideal case, there is a big gap between the remnant polarization in HZO and the maximum charge density in $Al_2O_3$. This fact can also be generally applied to other FE/DE stack systems. Thus, the puzzle and confusion in the field is how can ferroelectric polarization switching happen in a FE/DE stack without sufficient charge balance? Such charge difference can only be explained by introducing the leakage current and interfacial charges. The impact of leakage current in FE/DE bi-layer was previously reported, and the interfacial charging is believed to be important in the ferroelectric switching process by a thermodynamic free energy model.[10] In FE hafnium oxide systems, the discussions are mostly focused on the impact of leakage current on the negative capacitance effect.[15-17,25] The hysteresis-free NC effect in HZO/$Al_2O_3$ are reported by fast pulse measurement, the impacts of leakage current and charge trapping are minimized because of the fast pulses.[15-17,26]

In this work, we provide a simple understanding on the ferroelectric polarization switching process in FE/DE stack by introducing the leakage current through the thin DE layer and only considering the electrostatics. The ferroelectric polarization switching of HZO in the HZO/$Al_2O_3$ FE/DE stack is investigated systematically by capacitance-voltage (C-V) and polarization-voltage (P-V) measurements. The thickness of dielectric layer is found to have determinant impact on the ferroelectric polarization switching of ferroelectric HZO. The suppression of ferroelectricity is observed with thick dielectric layer. In the gate stacks with thin dielectric layers, a full polarization switching of the ferroelectric layer is found possible by the proposed leakage-current-assist mechanism through the ultrathin dielectric layer. It is confirmed that the charge needed for ferroelectric polarization switching comes from the leakage current through the thin dielectric layer. Without such leakage current to realize the charge balance, the FE HZO cannot be fully polarized.

**Experimental**

Fig. 1 shows the experimental device structures. Four types of capacitor structures are used in this work: (a) TiN/$Al_2O_3$/TiN (type A), (b) TiN/HZO/TiN (type B), (c) TiN/$Al_2O_3$/HZO/TiN (type C), and (d) TiN/$Al_2O_3$/TiN/HZO/TiN (type D). The device fabrication process started with the standard solvent cleaning of heavily p-doped Si substrates (resistivity < 0.005 Ω·cm). TiN was deposited by atomic layer deposition (ALD) at 250 °C, using $[(CH_3)_2N]_4Ti$ (TDMAT, heated up to 60 °C) and $NH_3$ as the Ti and N precursors, respectively. All TiN layers are metallic and 30 nm thick. $Hf_{1-x}Zr_xO_2$ film was deposited by ALD at 200 °C, using $[(CH_3)_2N]_4Hf$ (TDMAHf, heated up to 60 °C), $[(CH_3)_2N]_4Zr$ (TDMAZr, heated up to 75 °C), and $H_2O$ as the Hf, Zr, and O precursors, respectively. The $Hf_{1-x}Zr_xO_2$ film with x = 0.5 was achieved by controlling $HfO_2$:$ZrO_2$ cycle ratio of 1:1. The ALD deposition of TiN and HZO were in two separated ALD chambers to avoid cross-contamination. The two ALD chambers are connected externally by Ar environment in a glove box to avoid the environmental contamination. After the deposition of type A-D structures, the samples were annealed at 500 °C in $N_2$ environment for 1 min by rapid thermal annealing. Then, Ti/Au top electrodes were fabricated by photo-lithography, e-beam evaporation and lift-off process (capacitor area=5024 $\mu m^2$). $CF_4$/Ar dry etching was done to remove top TiN layer for device isolation for type A-C capacitors. For type D capacitors, $BCl_3$/Ar dry etching was used to remove the top $Al_2O_3$ layer and $CF_4$/Ar dry etching was used to remove the top and middle TiN layers. All electrical measurements were done at room temperature in a cascade summit probe station. C-V measurement was performed using an Agilent E4980A LCR meter and P-V measurement was carried out using a Radiant RT66C ferroelectric tester.

**Results and Discussion**

Fig. 2(a) shows the C-V measurement of a type A capacitor with 20 nm $Al_2O_3$, from 1 kHz to 1 MHz. Fig. 2(b) shows the P-V measurement of the same type A capacitor, showing a linear dielectric characteristic. Both measurements (small signal C-V, dP/dV in P-V) give consistent capacitance values for the type A dielectric capacitor with a capacitance of ~0.33 $\mu F/cm^2$ and a corresponding dielectric constant of ~8. Fig. 3(a) shows the C-V measurement of a type B capacitor with 20 nm HZO, from 1 kHz to 1 MHz. The C-V measurement of a type B capacitor shows signature two peaks in the C-V hysteresis loop as the ferroelectric characteristics. The different capacitances at different voltages in C-V are attributed to the different dielectric constant due to the difference in atomic structures in different ferroelectric polarization states. The corresponding dielectric constants are calculated, as also shown in the right axis. Fig. 3(b) shows the P-V measurement of the same type B capacitor, showing a ferroelectric hysteresis loop.

Fig. 4(a) shows the C-V measurements of type C capacitors with 20 nm HZO and $Al_2O_3$ from 0 nm to 20 nm, measured at 10 kHz. The capacitances of type C capacitors decrease with thicker $Al_2O_3$ as expected. The signature two capacitance peaks due to ferroelectricity in the C-V hysteresis loop decrease and eventually disappear in 20 nm HZO/20 nm $Al_2O_3$ stack, suggesting the reduction of ferroelectricity in thick DE layer and FE layer stack. This feature is even more clearly presented in Fig. 4(b) which shows the P-V measurements of type C capacitors with 20 nm HZO and $Al_2O_3$ from 4 nm to 20 nm. The applied voltage ranges are maximized in P-V measurement before the leakage current has essential impacts. The significant decrease of remnant polarization in P-V hysteresis loops is clearly observed with thicker DE layers. The C-V measurements and P-V measurements consistently confirm that thick DE layer can suppress the ferroelectricity in FE/DE stack. Fig. 4(c) shows the P-V characteristics of a FE/DE capacitor with 20 nm HZO and 6 nm $Al_2O_3$, measured at different voltage sweep ranges. The coercive voltage and remnant polarization are found to be dependent on the sweep voltage range.

To further understand the physics behind the DE layer thickness dependence on the ferroelectricity of FE/DE stack, a theoretical analysis is provided, as shown in Fig. 5. To understand the dynamic process of ferroelectric switching, this process is plotted by a two-step process: before ferroelectric polarization switching and after ferroelectric polarization switching. As is well-known, the ferroelectric polarization switching is atom re-position within the unit cell, so it is always slower than the electron re-distribution. Thus, the two-step assumption is valid. For simplicity, it is assumed the FE layer has a dielectric constant of $\epsilon_{FE}$ (without considering ferroelectric polarization) and a thickness of $t_{FE}$; the DE layer has a dielectric constant of $\epsilon_{DE}$ and a thickness of $t_{DE}$. It also assumes that the FE layer has equal number of polarization up states and polarization down states in the virgin state before the measurement so the net polarization is zero, as in Fig. 5(a). This situation is similar to two high-k dielectric stack. We define $V_{TOT}$ to be the voltage applied to the FE/DE stack, $V_{DE}$ to be the voltage across the DE layer and $V_{FE}$ to be the voltage across the FE layer. Therefore, before the ferroelectric polarization switching, the voltages across the DE layer and FE layer are

$$\boldsymbol{V_{DE,init}} = \frac{V_{TOT}\epsilon_{FE}t_{DE}}{\epsilon_{DE}t_{FE}+\epsilon_{FE}t_{DE}} \quad (1)$$

$$\boldsymbol{V_{FE,init}} = \frac{V_{TOT}\epsilon_{DE}t_{FE}}{\epsilon_{DE}t_{FE}+\epsilon_{FE}t_{DE}} \quad (2)$$

There are totally three different cases according to the different DE thickness. Here, we first assume the leakage current is zero and then discuss the impact of leakage current. Firstly, if $t_{DE}$ is very thick, then $V_{FE,init}$ can be sufficiently small so that it is smaller than the coercive voltage ($V_c$) of the FE layer, according to eqn. (2). Thus, no polarization switching can happen. So, the C-V and P-E characteristics behave like a linear dielectric insulator. Secondly, if $V_{FE,init} > V_c$ but $t_{DE}$ is sufficiently thick, the FE layer cannot be fully polarized. As the polarization switching happens, $V_{DE}$ increases until $V_{FE}$ reaches $V_c$ and the polarization process cannot continue. In this case, the total charge in FE layer ($Q_{FE}$) can be approximated as $\epsilon_{DE}(V_{TOT}-V_c)/t_{DE}$, where we have ferroelectric polarization charge ($P_{FE}$) < $P_r$ and $V_{DE,final}=Q_{FE}/(\epsilon_{DE}/t_{DE})$. Note that $P_{FE} \cong Q_{FE}=Q_{DE}$ if the $P_{FE}$ is significantly larger than the dielectric charge. Such assumption is made for the simplicity of qualitative discussion and does not affect the conclusion. The numerical simulation including the difference of $P_{FE}$ and $Q_{FE}$ gives the same conclusion. Thirdly, if the DE layer is thin enough, so that the second criterion does not meet anymore, we can have the FE layer fully polarized. So $V_{DE}$ can be estimated as $V_{DE,final}=P_r/(\epsilon_{DE}/t_{DE})$. It is clear that if $Q_{FE}$ is larger than the maximum charge density in DE layer, $V_{DE,final}$ will be larger than the breakdown voltage ($V_{BD}$) of the DE layer, which of course cannot happen. What is really happening in this process (if $V_{DE,final} > V_{BD}$) is when $V_{DE}$ rises from $V_{DE,init}$ to $V_{DE,final}$, the DE layer first becomes leaky and these leakage charges will balance the ferroelectric polarization charges so that $V_{DE}$ cannot reach $V_{BD}$. Thus, all the ferroelectric polarization charges are balanced by the charges from leakage current instead of the charge in DE layer. Therefore, in thin DE limit, the ferroelectric polarization switching process is a leakage-current-assist process. At the extremely leaky limit, it becomes almost as metal-FE-metal structure. Here's a summary of all three cases,

Case 1: thick DE limit, no polarization switching

$$\boldsymbol{V_{FE,init} < V_c} \qquad (3)$$

Case 2: moderate DE thickness, partial switching (by dielectric charge or leakage)

$$\boldsymbol{V_{FE,init} > V_c \ and \ P_r > \epsilon_{DE}(V_{TOT} - V_c)/t_{DE}} \quad (4)$$

$$\boldsymbol{P_{FE} = \epsilon_{DE}(V_{TOT} - V_c)/t_{DE}} \quad (5)$$

$$\boldsymbol{V_{DE,final} = P_{FE}/\frac{\epsilon_{DE}}{t_{DE}}} \quad (6)$$

Case 3: ultra-thin DE limit, leakage-current-assist switching

$$\boldsymbol{P_r < \epsilon_{DE}(V_{TOT} - V_c)/t_{DE}} \quad (7)$$

$$\boldsymbol{P_{FE} = P_r} \quad (8)$$

$$\boldsymbol{V_{DE,final} = P_r/\frac{\epsilon_{DE}}{t_{DE}}} \quad (9)$$

To calculate the leakage-current-assistant switching process as in case 3, a theoretical model is developed. Fig. 5 shows the model of FE/DE stack and the charge distribution upon the application of a positive external voltage ($V_{TOT}$). If there is no leakage current, charge in the DE layer ($Q_{DE}$) is always the same as the charge in the FE layer ($Q_{FE}$). But as discussed above, this no leakage current assumption is not valid since $Q_{FE}$ can be much larger than the maximum $Q_{DE}$ at DE breakdown. Therefore, the leakage current through the DE layer is unavoidable. Here, $E_{effect}$ is defined as a critical electric field. For simplicity, $E_{effect}$ is assumed to be a constant without thickness dependence. There is negligible leakage current below the $E_{effect}$ and above the $E_{effect}$ the leakage current exists. The charge carried by the leakage current will be trapped at the FE/DE interface as $Q_{it}$. In the equilibrium condition, there is no charge transfer process with zero current. As a result, the electric field in the DE layer will be pinned at the $E_{effect}$, so that,

$$\boldsymbol{Q_{DE} = \epsilon_{DE}E_{effect}} \quad (10)$$

The charge balance equation becomes

$$\boldsymbol{Q_{FE} = Q_{DE} + Q_{it}} \quad (11)$$

As we can see, it is critical to have enough $Q_{it}$ from leakage current to obtain a high $Q_{FE}$. So, the polarization switching process must be leakage-assist-switching. Eqns. (10) and (11) are the key formulas in the simulation of P-V hysteresis loop of FE/DE stack.

Fig. 6(a) shows the simulation of the minor loops of the FE HZO capacitor based on numerical fitting to the experimental P-E curve. If there is not enough charge from $Q_{it}$ and $Q_{DE}$, the FE layer always exhibits a minor loop with less $P_r$. Fig. 6(b) shows the simulation of the P-V hysteresis loops in a 6 nm $Al_2O_3$/20 nm HZO capacitor at different voltage sweep ranges, assuming no leakage current. Fig. 6(c) shows the simulation of the same structure but using the leakage-assist-switching model with leakage current from DE layer. It is obvious that a significantly larger $P_r$ is obtained. The experimental P-V hysteresis loops in a 6 nm $Al_2O_3$/20 nm HZO capacitor (Fig. 4(c)) with certain level of unavoidable leakage current match well with the leakage-assist-switching model presented in Fig. 6(c).

The thickness of DE layer, thus the leakage current, has a significant impact on the ferroelectricity of FE/DE stack as studied experimentally. The leakage-assist-switching model can also simulate the thickness-dependent behavior. Fig. 7(a) shows the simulation of P-V hysteresis loop of FE/DE stack with 20 nm HZO and different $Al_2O_3$ thicknesses, assuming no leakage current. Fig. 7(b) shows the simulation of the same structure but using the leakage-assist-switching model. The specific voltage sweep ranges are selected according the experimental results, as shown in Fig. 4(b). Modeling and experiments are in great agreement. It also successfully predicts the loss of $P_r$ in FE/DE stack with thick DE layer, because of the voltage division and no leakage current. Fig. 7(c) shows the $P_r$ versus thickness comparison in w/o leakage model and w/ leakage model with different $E_{effect}$, suggesting FE/DE stack with a leakier DE layer can have larger $P_r$. Fig. 7(d) compares the experimental results on $P_r$ vs. $Al_2O_3$ thickness with simulation results, showing the leakage-assist-switching model matches well with experimental results in terms of thickness dependence. The above results provide new insights to understand the large amount of reported experimental results in NC-FETs usually with FE/DE stacks. FE and FE/DE are fundamentally different in terms of coercive field, $P_r$, switch speed, and many others.

The impact of internal metal is studied by comparing type C and type D capacitors (see supplementary section 1). The FE/DE stacks with internal metal and without internal metal are physically very different. If the internal metal gate becomes the externally connected metal wires or the internal metal gate is physically connected to measurement equipment, the required balanced charges can be provided even externally. All these facts are extremely important to understand and interpret the experimental observation related to Fe-FETs and NC-FETs. The so-called interfacial coupling effect or capacitance enhancement was observed in the previous reports[19,27-30]. This interfacial coupling effect can improve the equivalent oxide thickness of FE/DE gate stack, which is also observed in the HZO/$Al_2O_3$ FE/DE stack as shown in supplementary section 2. The above understanding of ferroelectric switching process helps to just apply ferroelectric polarization switching (sometimes calls transient negative capacitance effect) to explain the DC enhancement of ferroelectric-gated transistors without invoking QSNC concept (see supplementary section 3). The operation speed of such transistors could be eventually limited by the ferroelectric switching speed[31], which needs to be thoroughly investigated still.

**Conclusion**

In summary, the ferroelectric polarization switching in FE HZO in the HZO/$Al_2O_3$ FE/DE stack is investigated systematically by C-V and P-V measurements. The thickness of dielectric layer is found to have determinant impact on the ferroelectric polarization switching of HZO. The suppression of ferroelectricity is observed with a thick linear dielectric layer. In the gate stacks with thin dielectric layers, a full polarization switching of the ferroelectric layer is found possible by the proposed leakage-current-assist mechanism through the ultrathin dielectric layer. The numerical simulation using the leakage-assist-switching model matches very well with the experimental results.

ASSOCIATED CONTENT

**Supporting Information**

Additional details for the impact of internal metal, interfacial coupling and the DC enhancement of Fe-FETs are in the supporting information.

AUTHOR INFORMATION

**Corresponding Author**

*E-mail: yep@purdue.edu

**Author Contributions**

P.D.Y. conceived the idea of FE/DE stack and supervised the experiments. M.S. and X.L. did the ALD deposition and device fabrication. M.S. and X.L. performed DC electrical measurements and analysis. M.S. did the numerical simulation. P.D.Y. and M.S. proposed the idea of DC enhancement on Fe-FETs. M.S. and P.D.Y. co-wrote the manuscript and all authors commented on it.

**Notes**

The authors declare no competing financial interest.

ACKNOWLEDGEMENTS

The authors would like to thank Muhammad Ashraful Alam and Suman Datta for valuable discussions. The work was supported in part by the Semiconductor Research Corporation (SRC) and DARPA.

**Figures**

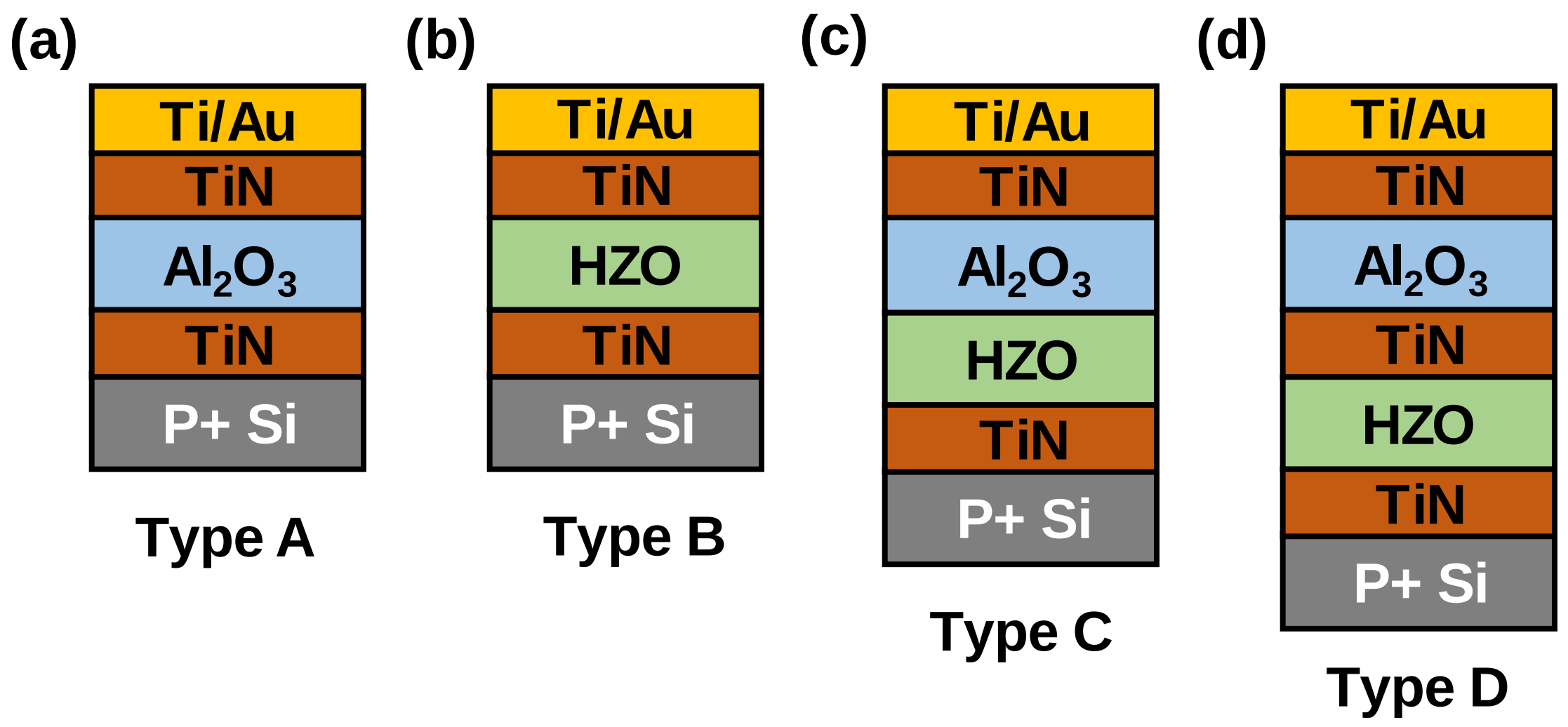

**Figure 1.** Capacitor structures used in this work (a) $Al_2O_3$ only (type A), (b) HZO only (type B), (c) $Al_2O_3$ and HZO stack without internal metal (type C), and (d) $Al_2O_3$ and HZO stack with TiN layer as the internal metal (type D).

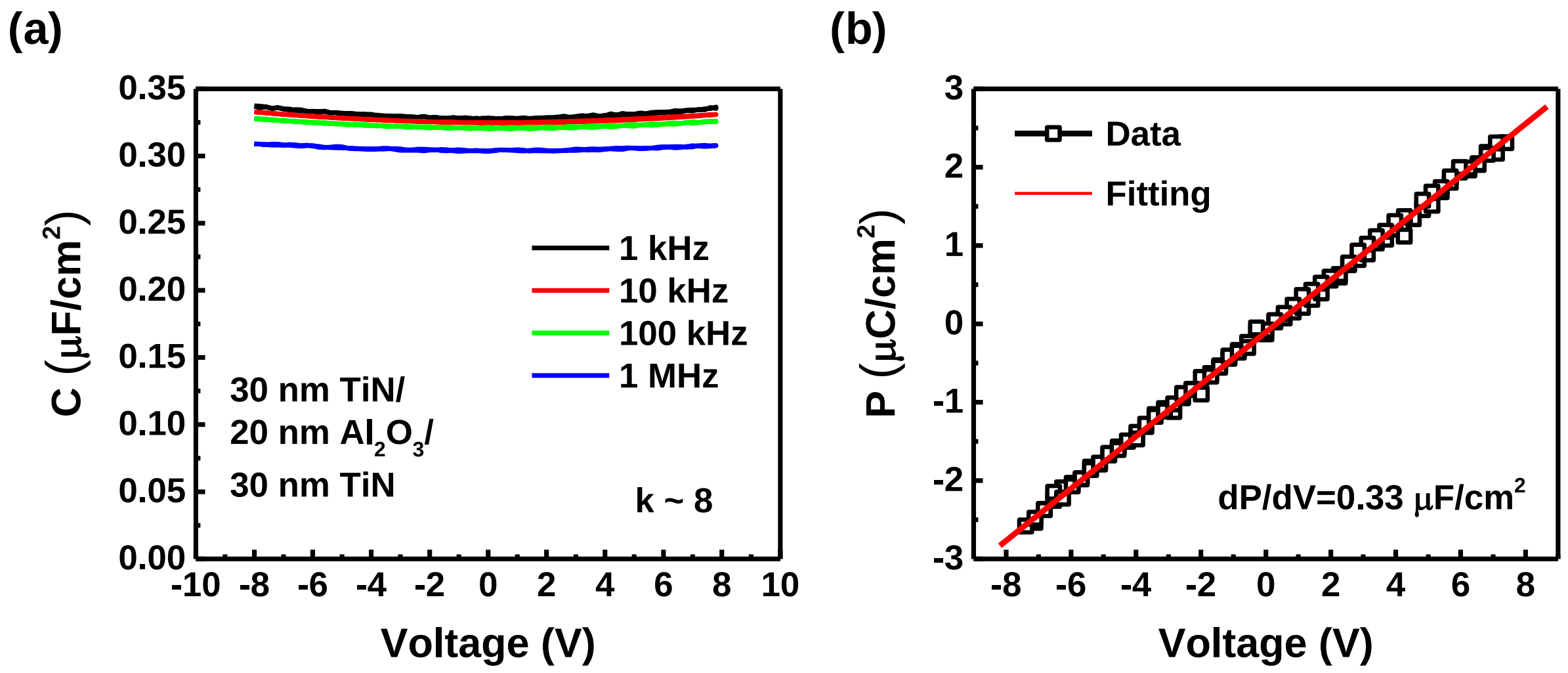

**Figure 2.** (a) C-V measurement and (b) P-V measurement on a type A capacitor with 20 nm $Al_2O_3$.

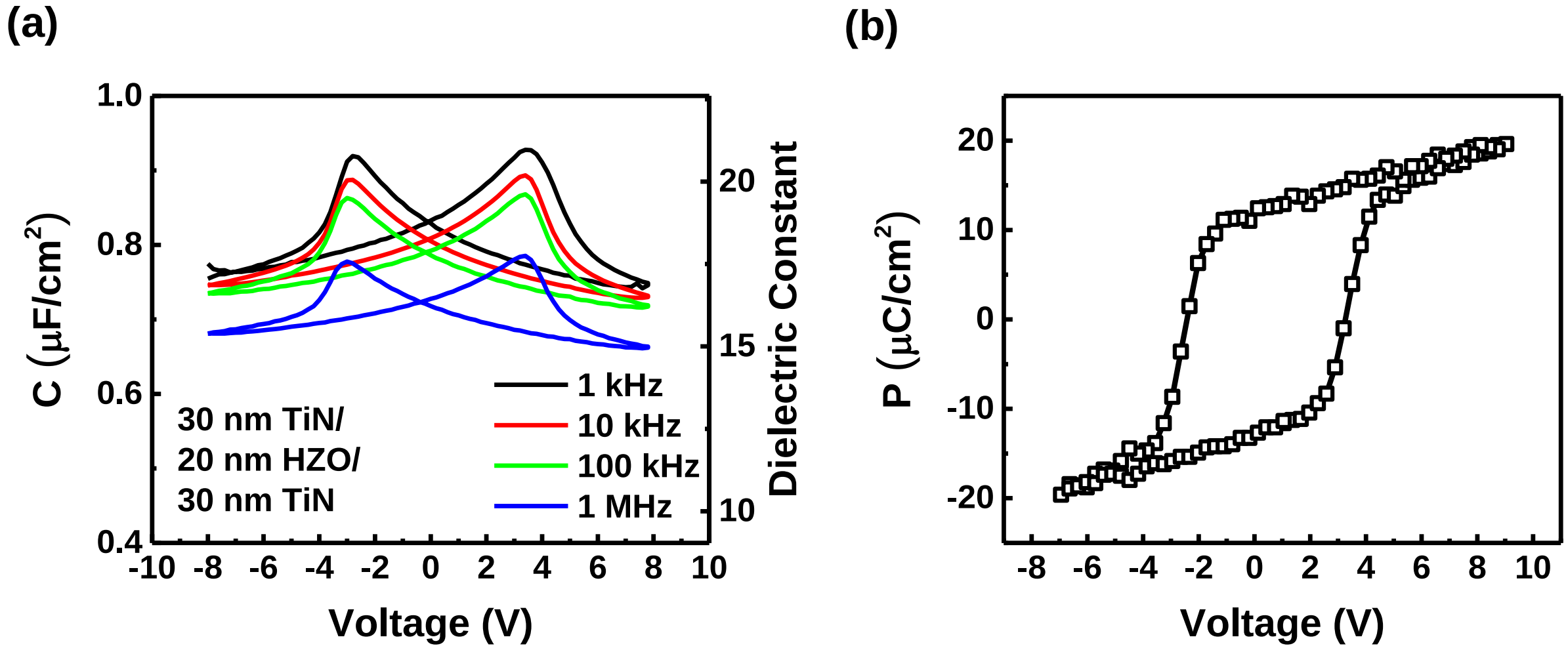

**Figure 3.** (a) C-V measurement and (b) P-V measurement of a type B capacitor with 20 nm HZO.

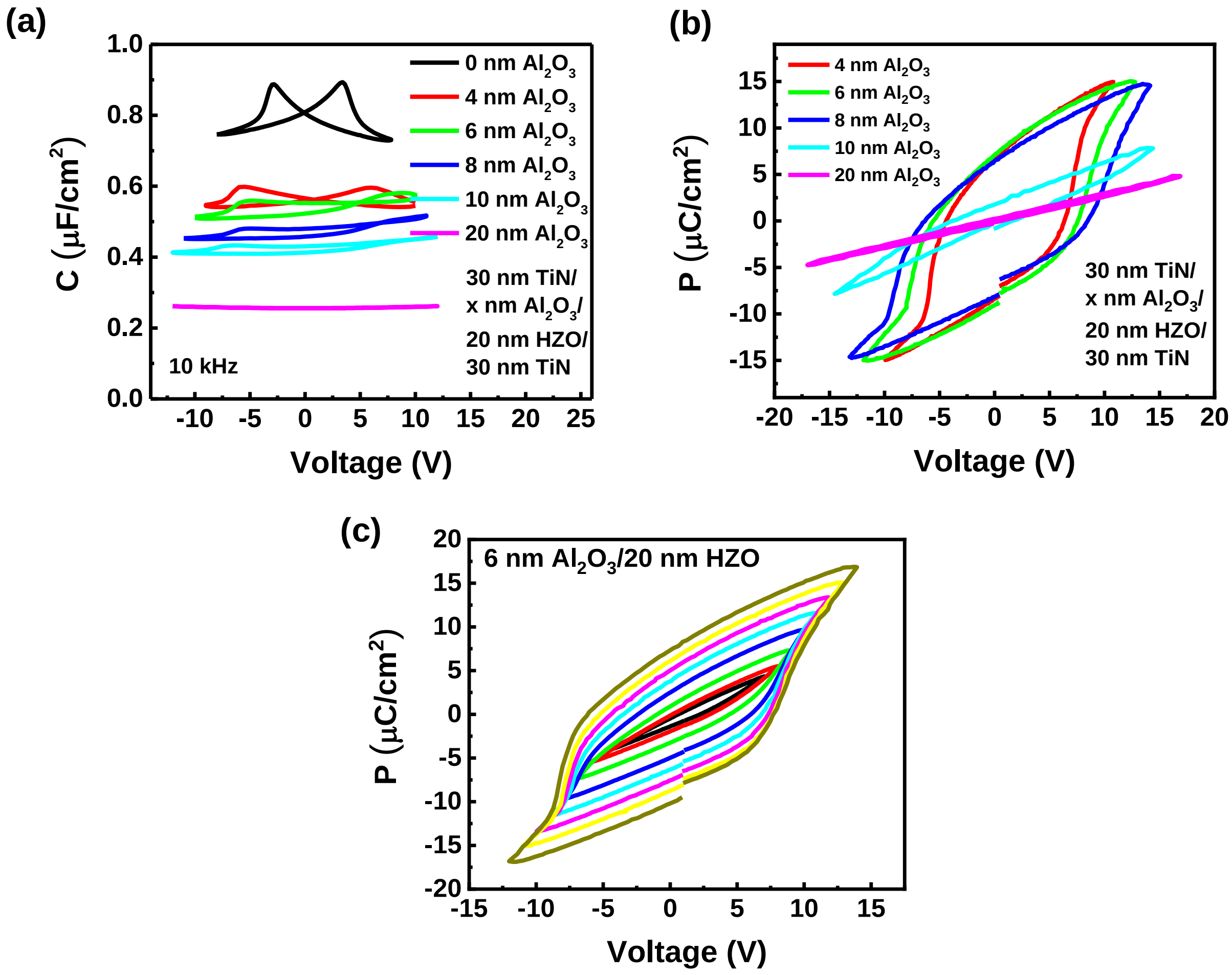

**Figure 4.** (a) C-V measurements and (b) P-V measurements on type C capacitors with different $Al_2O_3$ thickness and 20 nm HZO. (c) P-V measurements on a type C capacitor with 6 nm $Al_2O_3$ and 20 nm HZO at different voltage sweep ranges.

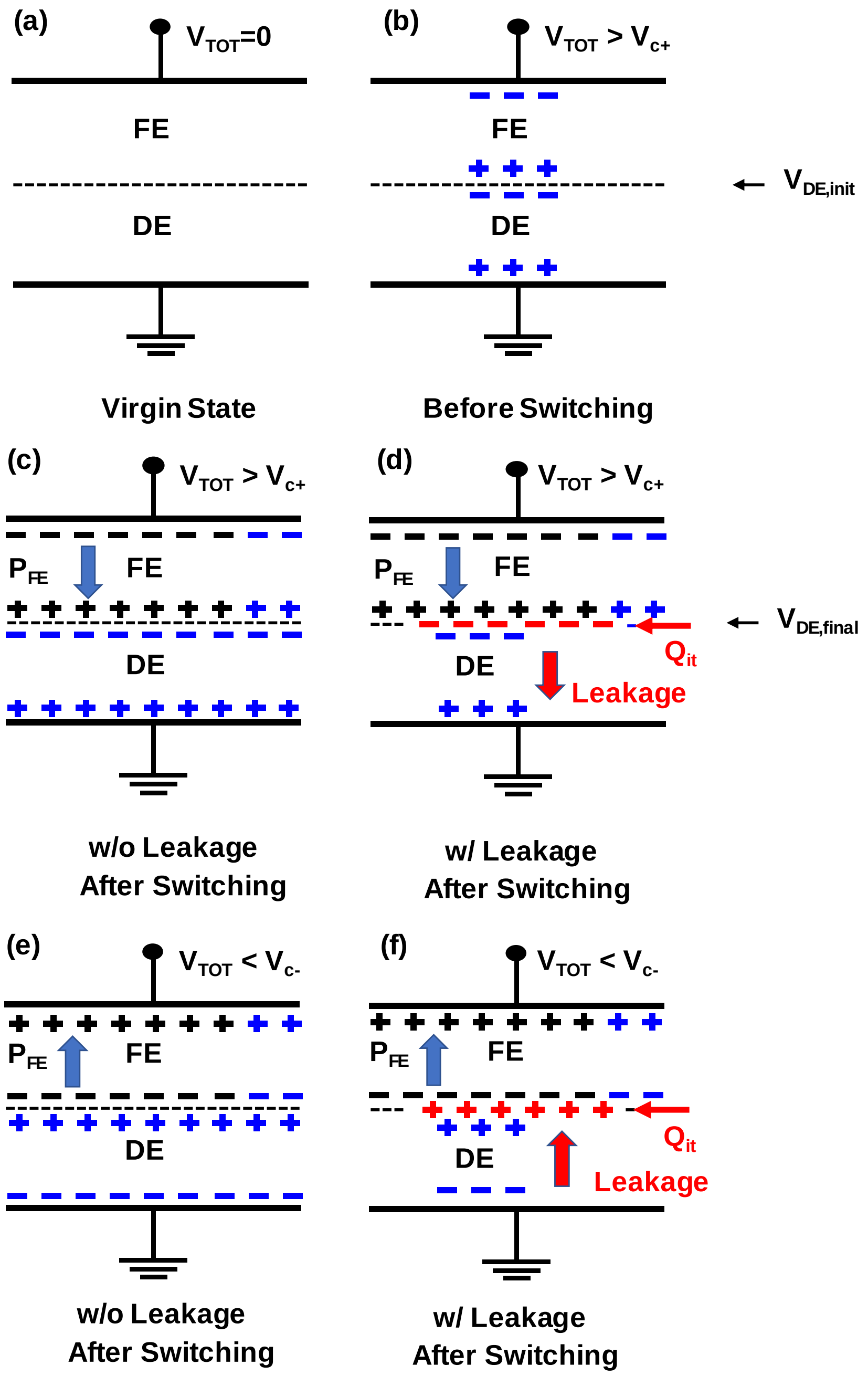

**Figure 5.** Illustration of charge distribution in FE/DE stack in the following conditions. (a) $V_{TOT}=0$ and FE layer is not polarized. (b) $V_{TOT}$ larger than positive coercive voltage $V_{c+}$ before polarization

switching. $V_{TOT}$ larger than positive coercive voltage $V_{c+}$ after polarization switching (c) assuming no leakage current though DE layer and (d) assuming the existence of leakage current though DE layer. $V_{TOT}$ less than negative coercive voltage $V_{c-}$ after polarization switching (e) assuming no leakage current though DE layer and (f) assuming the existence of leakage current though DE layer. Type of charges: black, ferroelectric polarization; blue, dielectric polarization; red, interfacial trapped charge.

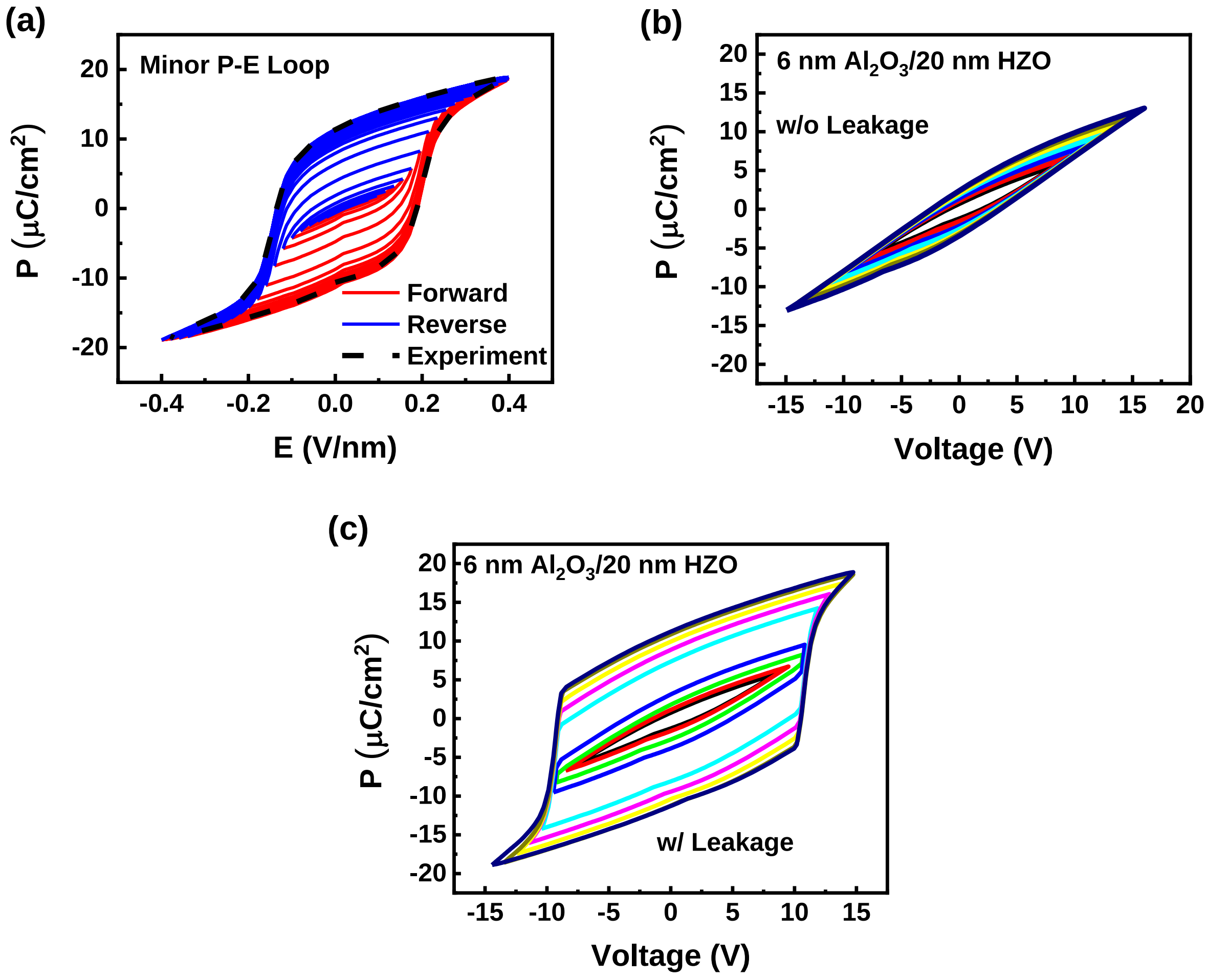

**Figure 6.** (a) Simulation of minor loops in a FE HZO capacitor. (b) Simulation w/o leakage current of P-V hysteresis loops in a 6 nm $Al_2O_3$/20 nm HZO capacitor at different voltage sweep ranges.

(c) Simulation w/ leakage current of P-V hysteresis loops in a 6 nm $Al_2O_3$/20 nm HZO capacitor at different voltage sweep ranges.

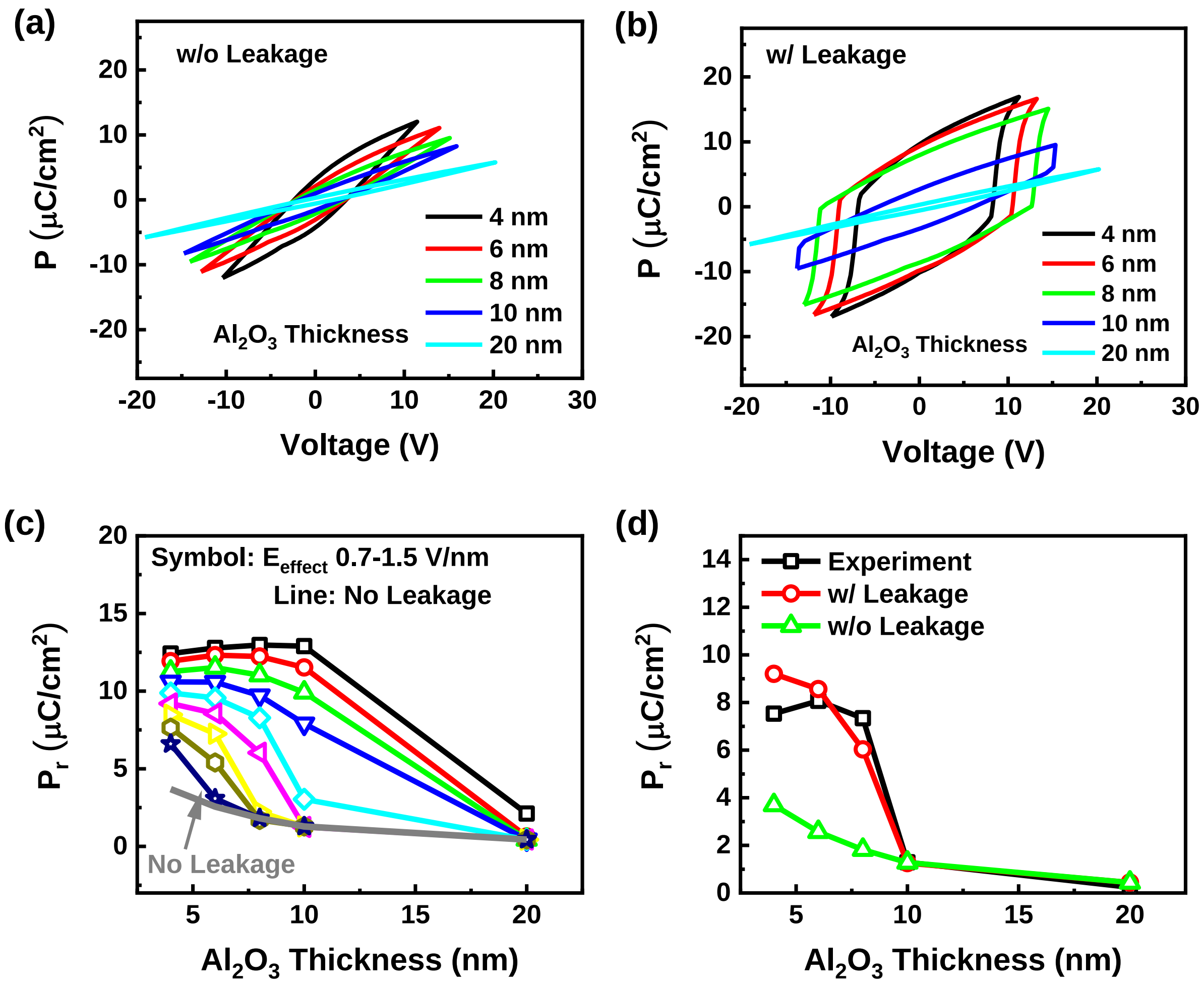

**Figure 7.** Simulation on (a) w/o leakage current and (b) w/ leakage current of P-V hysteresis loops of FE/DE capacitors with 20 nm HZO and different $Al_2O_3$ thicknesses. (c) Remnant polarization versus $Al_2O_3$ thickness on both w/o leakage current and w/ leakage current assumptions at different $E_{effect}$. (d) Comparison of experimental results on $P_r$ vs. $Al_2O_3$ thickness with simulation results, w/ and w/o leakage-assist-switching.

TOC

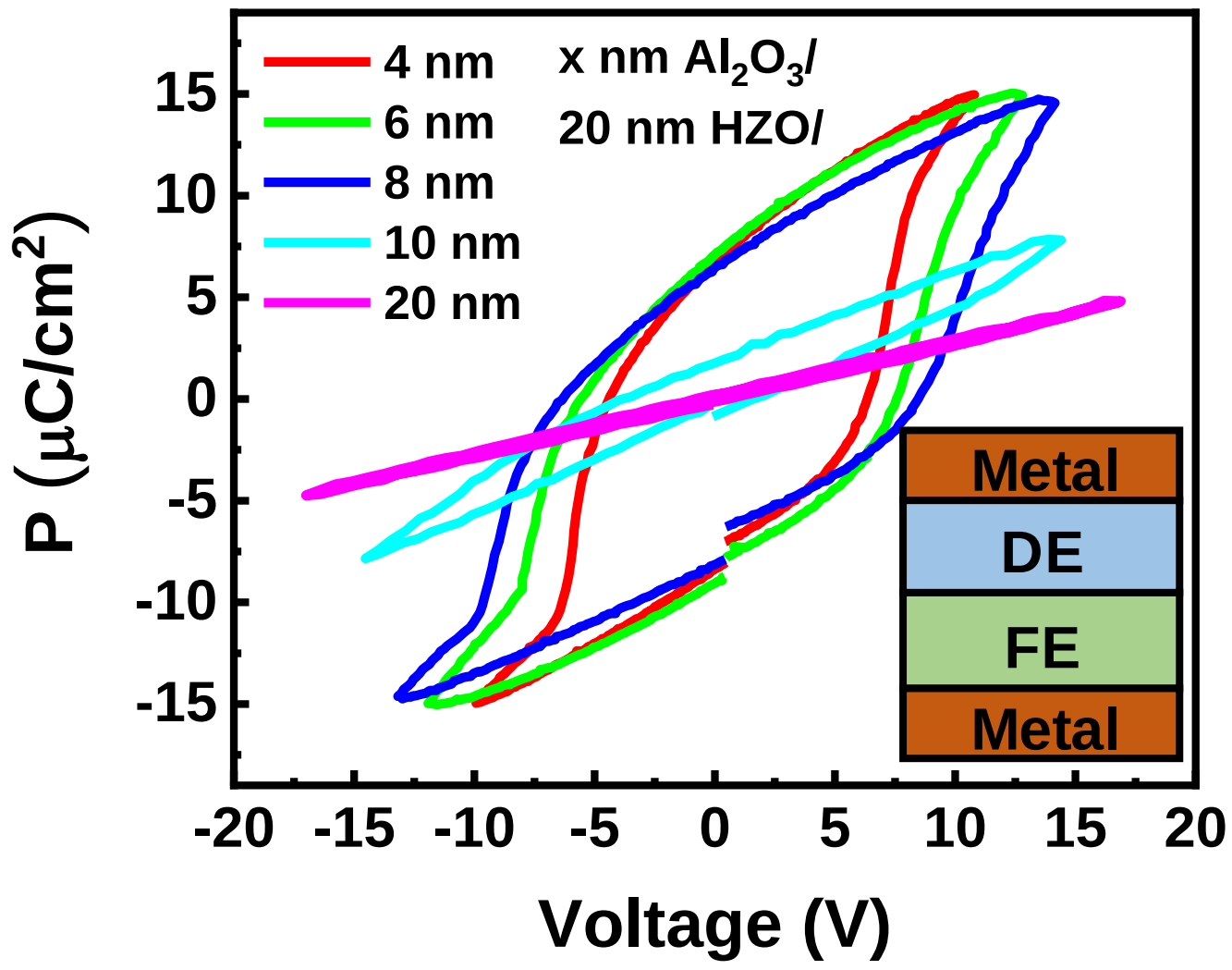
4 nm
6 nm
8 nm
10 nm
20 nm
x nm $Al_2O_3$/
20 nm HZO/
Metal
DE
FE
Metal
P (μC/cm²)
Voltage (V)
15
10
5
0
-5
-10
-15
-20 -15 -10 -5 0 5 10 15 20

Supplementary Information for:

# On the Ferroelectric Polarization Switching of Hafnium Zirconium Oxide in Ferroelectric/Dielectric Stack

Mengwei Si, Xiao Lyu, and Peide D. Ye*

*School of Electrical and Computer Engineering and Birck Nanotechnology Center, Purdue University, West Lafayette, Indiana 47907, United States*

* Address correspondence to: yep@purdue.edu (P.D.Y.)

## 1. Ferroelectric/Dielectric Stack with Internal Metal

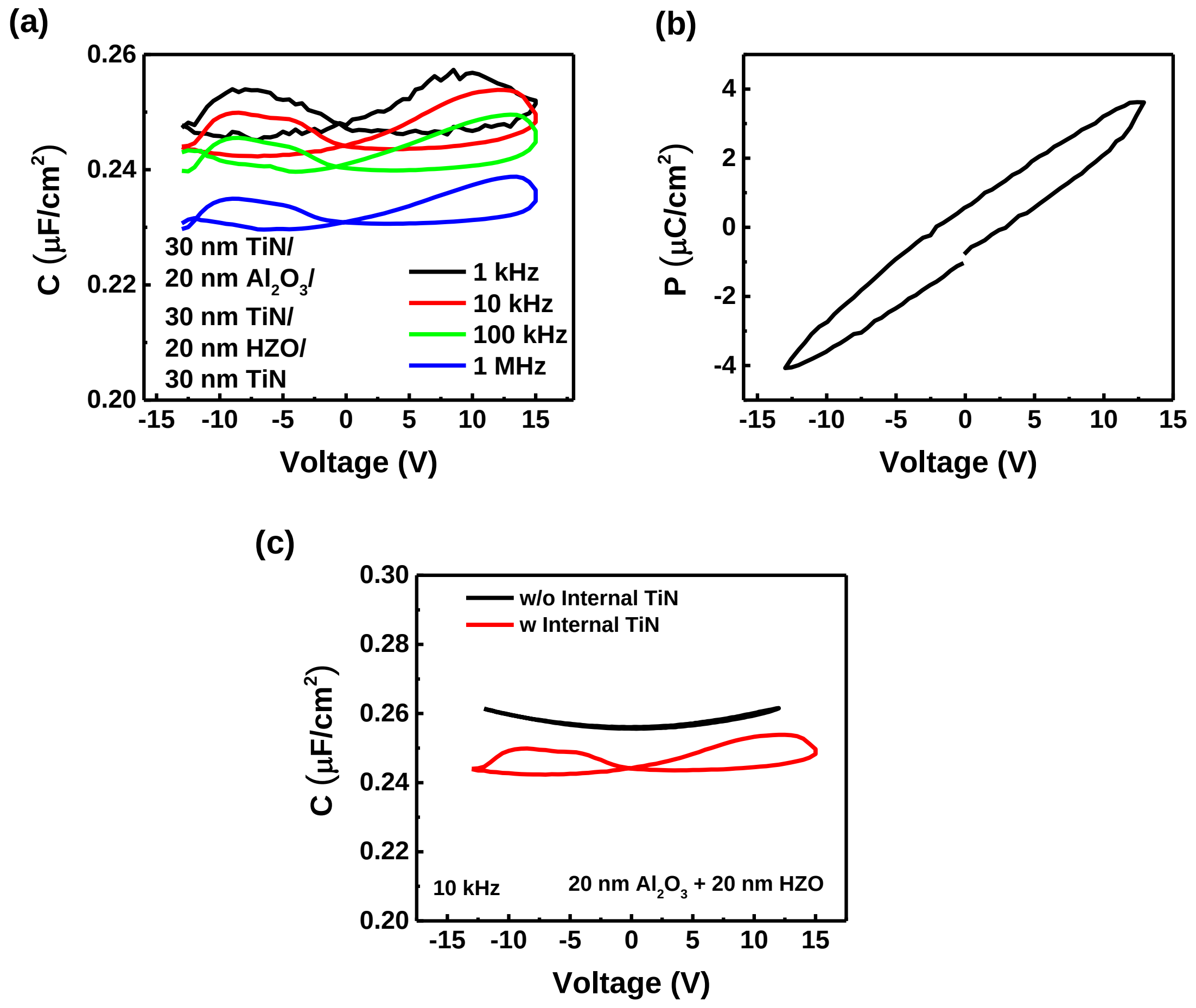

**Figure S1.** (a) C-V measurement and (b) P-V measurement on a type D capacitor with 20 nm $Al_2O_3$ and 20 nm HZO. (c) C-V measurements comparison of a type C and a type D capacitor with 20 nm $Al_2O_3$ and 20 nm HZO.

Fig. S1(a) shows the C-V measurement of a type D capacitor with 20 nm HZO, 20 nm $Al_2O_3$ and 30nm TiN in between, measured from 1 kHz to 1 MHz. The two signature capacitance peaks in the C-V hysteresis loop are observed. Fig. S1(b) shows the P-V measurement of a type D capacitor with 20 nm HZO and 20 nm $Al_2O_3$, where a weak ferroelectric hysteresis loop is achieved. Fig. S1(c) shows the comparison of C-V measurements of type C and type D capacitors with 20 nm HZO and 20 nm $Al_2O_3$. Although same thicknesses of HZO and $Al_2O_3$ are used, a type D capacitor with 30 nm TiN exhibits weak ferroelectricity in C-V and P-V characteristics, suggesting that the charge in the internal metal can assist the ferroelectric switching process, in great contrast to the result from a type C capacitor. Fig. S1(c) concludes the FE/DE stacks with internal metal and without internal metal are physically very different. If the internal metal gate becomes much larger than the capacitor area by design or it is externally connected to metal wires through vias or the internal metal gate is physically connected to the measurement equipment, the required balanced charges can be provided even externally. All these facts are extremely important to understand and interpret the experimental observation related to Fe-FETs and NC-FETs.

## 2. Interfacial Coupling in Ferroelectric/Dielectric Stack

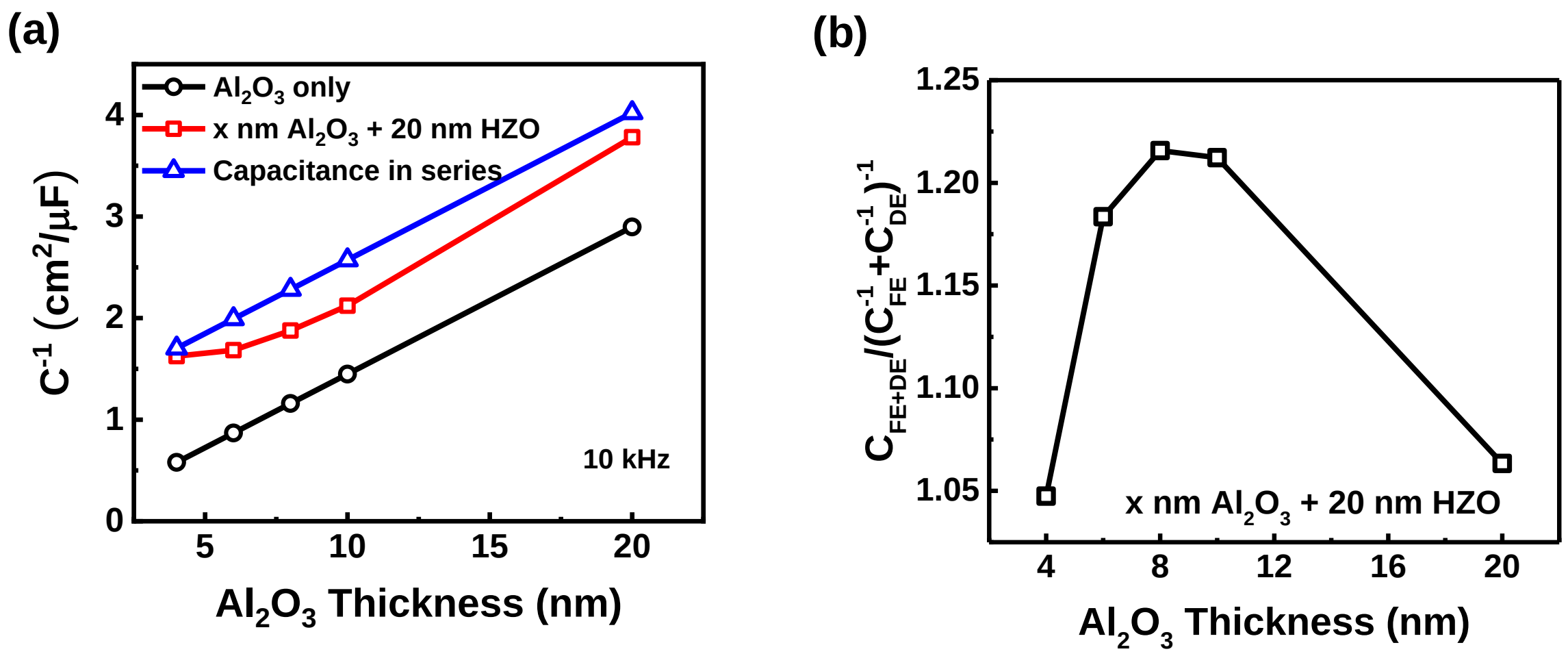

**Figure S2.** (a) Capacitance$^{-1}$ of type C capacitor versus $Al_2O_3$ thickness, by both experimental measurements (red squares) and calculated total capacitance (blue triangles). The calculated total capacitance in series is based on the experimental measurements of type A and type B capacitors, using the maximum capacitance in the measured C-V curves. The capacitance values of measured type A capacitors with different $Al_2O_3$ thicknesses are presented as black circles. (b) The ratio of experimental capacitances of type C capacitor over capacitance in series versus different $Al_2O_3$ thickness.

Fig. S2(a) shows the capacitance$^{-1}$ versus $Al_2O_3$ thickness characteristics of three types of capacitors, $Al_2O_3$ only (type A), $Al_2O_3$/20 nm HZO stack (type C) and the capacitance value of measured $Al_2O_3$ (type A) and HZO (type B) capacitors in series. Experimentally, capacitance of type C capacitor is lower than type A capacitor with same $Al_2O_3$ thickness. No obvious QSNC effect is observed in HZO material system. But a capacitance enhancement of type C capacitor is observed to be larger than the capacitance value in series, as shown in Fig. S2(b). Over 20% capacitance enhancement is observed with 8 and 10 nm $Al_2O_3$/20 nm HZO stack. This result shows from another aspect that the FE/DE stack with internal metal and without internal metal are physically very different. It demonstrates the existence of interfacial coupling[1-6] between the $Al_2O_3$ layer and HZO layer. This interfacial coupling effect can improve the equivalent oxide thickness of FE/DE gate stack. The static capacitance enhancement by negative capacitance effect ($C_{TOT}$ > $C_{DE}$) is not directly achieved in this slow measurement. The intrinsic quasi-static negative capacitance phenomenon might be masked by the charge trapping and de-trapping.[10] So it is not conclusive to claim the existence of negative capacitance or not in this work using HZO as the ferroelectric stack.

### 3. DC Enhancement in Fe-FET

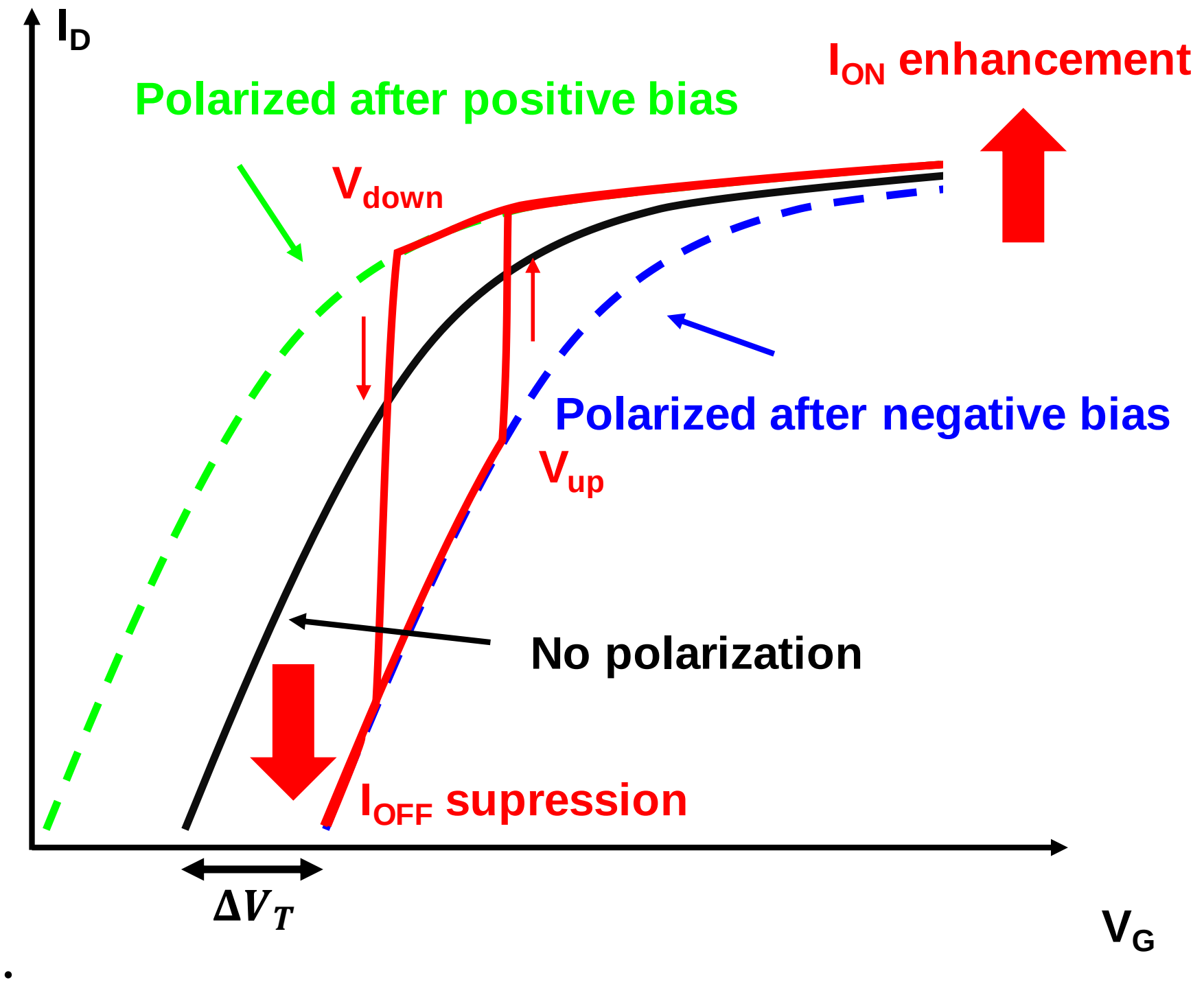

**Figure S3.** (a) Illustration of DC enhancement of a ferroelectric-gated FET.

It is clear that ferroelectric polarization switching can lead to the sub-60 mV/dec subthreshold slope (SS) in ferroelectric-gated transistors. But hysteresis in transfer characteristics is unavoidable, if not considering charge trapping process. Note that the concept of transient NC effect in Fe-FETs[7-9] is fundamentally different from the concept of QSNC effect in NC-FETs. But it is unclear whether performance benefit is achievable or not with a hysteretic and sub-60 mV/dec device. Here, the authors want to emphasize that ferroelectric polarization switching and polarization charge in Fe-FETs can offer DC enhancement ($I_{OFF}$ reduction and $I_{ON}$ enhancement simultaneously), with the existence of manageable hysteresis and without incorporating QSNC explanation. All transient effects of ferroelectric dynamic polarization switching[10–15] are negligible in DC condition discussed in this work. Meanwhile, the hysteresis might not have serious impact in logic circuits if it is controlled in between zero voltage and half of the supply voltage ($V_{DD}$).

Thus, this work addresses an important fact that ferroelectric field-effect transistors can offer DC enhancement from the perspective of ferroelectric polarization switching only. The potential of using ferroelectric-gated transistor for low-power logic applications is limited by the speed of the devices.

In a CMOS logic circuit, a lower off-state current ($I_{OFF}$) and higher on-state current ($I_{ON}$) is preferred. The DC enhancement here is defined as at the same $I_{OFF}$ and a given supply voltage ($V_{DD}$), the transistor can have higher $I_{ON}$. Whether a small hysteresis exists or not is not important if lower $I_{OFF}$ and higher $I_{ON}$ can be achieved simultaneously. For circuit applications, hysteresis window of the devices should be controlled less than half of the $V_{DD}$. As shown in Fig. S3, the black line is the transfer characteristics of the baseline FET without ferroelectric polarization. If a high gate voltage is applied, the transfer curve shifts to the left as the green curve. If a low gate voltage is applied, the transfer curve shifts to the right as the blue curve. The amount of threshold voltage shift ($\Delta V_T$) is determined by the remnant polarization, the capacitance of dielectric layer ($C_{DE}$) and the ratio of ferroelectric capacitor area ($A_{FE}$) and the dielectric capacitor area ($A_{DE}$) (assuming the existence of an internal metal layer). Note that the conclusion is still valid in FE/DE stack without internal metal, but the area ratio of $A_{FE}$ and $A_{DE}$ becomes one. The transfer characteristics of the Fe-FET switches between the polarization up and polarization down transfer curves and the switching voltages ($V_{up}$, $V_{down}$) are determined by the coercive voltages. The coercive voltages can be tuned by the thickness of the FE layer, so $V_{up}$ and $V_{down}$ can be tuned accordingly. Therefore, if we plot the full bi-directional transfer characteristics, as the red line in Fig. S3, a reduction in $I_{OFF}$ and an enhancement in $I_{ON}$ are achieved simultaneously. This exactly shows the DC enhancement can be achieved using a Fe-FET structure. The difficulty in realization of such performance is that the $P_r$ in conventional ferroelectric insulator material is so high that the hysteresis window become too large for logic applications. However, by using a DE layer for capacitance matching and using an internal metal gate to modulate the area ratio of $A_{FE}$ and $A_{DE}$ if it is needed, we can effectively reduce the hysteresis window, achieve DC enhancement in Fe-FET.

Such experimental structure and experimental results were already reported in our previous publication with $A_{FE}/A_{DE}$ ~ 100, as shown in Ref. 16. It is a $MoS_2$ ferroelectric-gate FET with internal metal gate structure. Subthreshold slope (SS) of 37.6 mV/dec in forward sweep and SS of 42.2 mV/dec in reverse sweep are achieved. More importantly, a clear $I_{ON}$ enhancement is achieved with same $I_{OFF}$ so that this is an obvious DC enhancement. From the perspective of ferroelectric polarization switching, such DC enhancement can be explained without invoking QSNC concept.